\documentclass[12pt]{iopart}
\usepackage{iopams}
\usepackage{graphicx}% Include figure files
\usepackage{dcolumn}% Align table columns on decimal point
\usepackage{bm}% bold math
\usepackage{mathbbol}
\usepackage{axodraw}
\usepackage{epsf}
\usepackage{mathrsfs}
\usepackage{slashed}
\usepackage{stmaryrd}
\newcommand{\be}{\begin{equation}}
\newcommand{\ee}{\end{equation}}
\newcommand{\bea}{\begin{eqnarray}}
\newcommand{\eea}{\end{eqnarray}}
\newcommand{\bml}{\begin{mathletters} \baselineskip 10pt}
\newcommand{\eml}{\baselineskip 12pt \end{mathletters}}
\newcommand{\nn}{\nonumber}

\newcommand{\bt}{\tilde{b}}

\def\lambdabar{\protect\@lambdabar}
\def\@lambdabar{%
\relax
\bgroup
\def\@tempa{\hbox{\raise.73\ht0
\hbox to0pt{\kern.2\wd0\vrule width.7\wd0
height.1pt depth.1pt\hss}\box0}}%
\mathchoice{\setbox0\hbox{$\displaystyle\lambda$}\@tempa}%
{\setbox0\hbox{$\textstyle\lambda$}\@tempa}%
{\setbox0\hbox{$\scriptstyle\lambda$}\@tempa}%
{\setbox0\hbox{$\scriptscriptstyle\lambda$}\@tempa}%
\egroup
}

\newcommand{\DL}{\delta \mathscr{L}}
\newcommand{\SCS}{\mathscr{S}}
\newcommand{\SCP}{\mathscr{P}}

\newcommand{\sfrac}[2]{{\textstyle \frac{#1}{#2}}}
\newcommand{\pad}[2]{\frac{\partial #1}{\partial #2}}

\newcommand{\vcb}[1]{\mbox{\bf #1}}

\newcommand{\vc}[1]{\mbox{\boldmath$#1$}}

\newcommand{\Ai}{\mbox{Ai}}

\newcommand{\Gi}{\mbox{Gi}}

\renewcommand{\Re}{\mbox{Re}}
\renewcommand{\Im}{\mbox{Im}}
\newcommand{\Ei}{\mbox{Ei}}

\begin{document}

\title{Large orders in strong-field QED}% Force line breaks with \\

\author{Thomas Heinzl}

\address{School of Mathematics and Statistics, University of
Plymouth\\
Drake Circus, Plymouth PL4 8AA, UK}

\ead{\mailto{theinzl@plymouth.ac.uk}}

\author{Oliver Schr{\"o}der}

\address{science + computing ag\\
Hagellocher Weg 73, D-72070 T{\"u}bingen, Germany}

\ead{\mailto{O.Schroeder@science-computing.de}}

%\date{\today}

\begin{abstract}
We address the issue of large-order expansions in strong-field
QED. Our approach is based on the one-loop effective action
encoded in the associated photon polarisation tensor. We
concentrate on the simple case of crossed fields aiming at
possible applications of high-power lasers to measure vacuum
birefringence. A simple next-to-leading order derivative expansion
reveals that the indices of refraction increase with frequency.
This signals normal dispersion in the small-frequency regime where
the derivative expansion makes sense. To gain information beyond
that regime we determine the factorial growth of the derivative
expansion coefficients evaluating the first 80 orders by means of
computer algebra. From this we can infer a nonperturbative
imaginary part for the indices of refraction indicating absorption
(pair production) as soon as energy and intensity become
(super)critical. These results compare favourably with an analytic
evaluation of the polarisation tensor asymptotics. Kramers-Kronig
relations finally allow for a nonperturbative definition of the
real parts as well and show that absorption goes hand in hand with
anomalous dispersion for sufficiently large frequencies and
fields.

\end{abstract}


%\submitto(\JPA)

\maketitle

\section{\label{sec:1}Introduction}

Recent years have seen a continuous progress in laser technology
leading to ever increasing values of power and intensity. This is
true for both optical lasers \cite{tajima:2002,hein:2004} and
systems based on free electrons like the DESY vacuum ultraviolet
free electron laser (VUV-FEL), a pilot system for the XFEL
radiating in the X-ray regime \cite{desy:2005}. High intensities
imply strong electromagnetic fields which approach magnitudes such
that vacuum polarisation effects may no longer be ignored even at
the comparatively low photon energies involved (1 eV ... 10 keV).
The theory describing these effects is strong-field quantum
electrodynamics (QED).

The most exciting and therefore best studied effect arises when
the ubiquitous virtual electron-positron pairs become real
(Schwinger pair production \cite{schwinger:1951}).  This happens
if the energy gain of an electron across a Compton wavelength
$\lambdabar_e$ equals its rest energy $m_e$ implying a critical
electric field
\be \label{ECRIT}
  E_c \equiv \frac{m_e^2}{e} \simeq 1.3 \times 10^{18} \,
  \mathrm{V/m} \; .
\ee
This also is often named after Schwinger although it has first
been obtained by Sauter upon solving the Dirac equation in a
homogeneous electric field \cite{sauter:1931}.

The physical situation to be analysed in this paper is as follows.
We assume a background field consisting of an intense, focused
laser beam of optical frequency $\Omega \simeq 1$~eV $\ll m_e$.
The associated gauge potential and field strengths are denoted
$A_\mu$ and $F_{\mu\nu}$, respectively. The laser beam
configuration is modeled by \textit{crossed fields} with electric
and magnetic fields constant, orthogonal and of the same
magnitude,
\be \label{CROSSED}
  E = B \equiv F = const \; , \quad \vc{E} \perp \vc{B} \; .
\ee
This configuration may be viewed as the zero-frequency limit of a
plane wave, $\Omega \to 0$. In covariant notation one may write
\be
  A_\mu = \sfrac{1}{2} F_{\mu\nu} x^\nu \; ,
\ee
with, for instance, $F_{01} = F = -F_{31}$ and all other entries
vanishing. Working with crossed fields leads to enormous
simplifications as will be seen below.

The crossed-field configuration will be probed by a weak
plane-wave field encoded in the potential $a_\mu$ and field
strength $f_{\mu \nu} = \partial_\mu a_\nu - \partial_\nu a_\mu$.
We denote the probe wave vector by
\be \label{INDEX}
  k = \omega (1, n \vcb{k}) \; ,
\ee
where $n \ge 1$ represents the index of refraction ($n=1$
\textit{in vacuo}) and the unit vector $\vcb{k}$ the direction of
propagation. Note that $n = 1/v$ with $v = \omega / |\vc{k}| \le
1$ being the phase velocity of the probe propagating through the
modified vacuum. This set-up has recently been suggested for an
experiment to measure vacuum birefringence for the first time
\cite{heinzl:2006}.

The corrections to pure Maxwell theory to leading order in the
fluctuation $a_\mu$ are given by the effective action
\cite{brezin:1970b}
\be \label{DELTA_S}
  \delta S \equiv \sfrac{1}{2} \int d^4 x \, d^4 y \; a_\mu (x)
  \Pi^{\mu \nu} (x,y; A) \, a_\nu(y) \; .
\ee
This is basically determined by the polarisation tensor (or photon
self-energy) $\Pi^{\mu\nu}$ in the presence of the background
field $A_\mu$.

In one-loop approximation the polarisation tensor is represented
by the Feynman diagram
\be \label{POLFEYN}
  \includegraphics[scale=1]{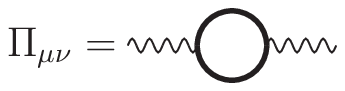}
\ee
where the heavy lines denote the fermion propagator dressed by the
background field $A$ (dashed lines below),
\be \label{DRESSEDPROP}
  \includegraphics[scale=1]{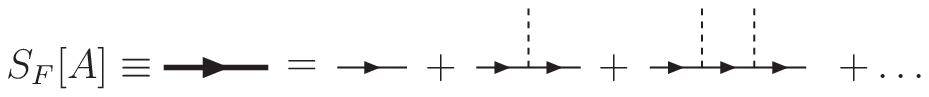}
\ee
Obviously, the first term on the right-hand side in
(\ref{DRESSEDPROP}) corresponds to vanishing background fields, $A
= 0$. In terms of the dressed propagator $S_F$ the polarisation
tensor (in momentum space) is given by the integral
\be
  \Pi^{\mu \nu} (k; A) = - i e^2 \tr \int \frac{d^4 p}{(2\pi)^4}
  \gamma^\mu S_F (p) \gamma^\nu S_F (p-k) \; ,
\ee
where `tr' denotes the Dirac trace.

The dressed propagator and the associated one-loop polarisation
tensor are known exactly for a few special backgrounds only (see
\cite{dittrich:2000} and \cite{dunne:2004} for overviews). In
particular, an exact solution is available for the crossed fields
(\ref{CROSSED}) due to Narozhnyi \cite{narozhnyi:1968} and Ritus
\cite{ritus:1972}. Their results may be written somewhat
symbolically in terms of a spectral decomposition,
\be \label{PI_SPEC1}
  \Pi^{\mu\nu} (k;A) = \sum_{i = 0, \pm}  \Pi_i (k;A) \,
  \epsilon_i^\mu \epsilon_i^\nu \; ,
\ee
with three orthogonal eigenvectors $\epsilon_i = \epsilon_i (k;A)$
satisfying $\epsilon_i \cdot k = 0$. The eigenvalues $\Pi_i$ are
given by somewhat complicated double integrals, the explicit form
of which will be given below. It is important to note that the
representation (\ref{PI_SPEC1}) is valid for arbitrary external
field strength and probe frequency. In other words, it contains
all orders in the background intensity and probe energy.

Let us briefly get an idea of the orders of magnitude involved. We
assume that our probe is provided by another laser, possibly of
high frequency, i.e.\ $\omega \gg \Omega$. For instance, if we
think of the projected DESY  XFEL as a probe with a projected wave
length of $\lambda = 0.1$ nm, hence $\omega \simeq$ 10 keV, we
expect a ratio $\omega/m_e \simeq 0.02 \ll 1$, that is \textit{low
energy} also for the probe. About the same frequencies can be
achieved for photons from a laser-based Thomson back-scattering
source \cite{schwoerer:2006}.

Assuming the background as an optical Petawatt laser focused down
to the diffraction limit the peak intensity will be
\cite{hein:2004,ringwald:2003}
\be \label{I_PETA}
  I = P/\Lambda^2 \pi \simeq 3 \times 10^{22} \; \mbox{W/cm}^2 \; ,
\ee
for $\Lambda = 2\pi /\Omega \simeq$ 1 $\mu$m and power $P$ = 1 PW.
Sauter's critical field (\ref{ECRIT}) corresponds to an intensity
of
\be \label{IC}
  I_c = E_c^2 = m_e^4/e^2 \simeq 4 \times 10^{29} \; \mbox{W/cm}^2 \; ,
\ee
still seven orders of magnitude larger than (\ref{I_PETA})
implying \textit{low intensity} for our background
laser\footnote{While the presently attainable fields are `weak'
compared to the critical field they are certainly very strong by
everyday standards. In particular they exceed \textit{static} lab
magnetic fields by orders of magnitude. This justifies the phrase
`strong-field QED' in the title.}. Hence for present technology we
have two primary small parameters \cite{heinzl:2006,affleck:1987},
namely
\bea
  \nu^2 &\equiv& \omega^2/m_e^2  \lesssim 4 \times 10^{-4} \;,
  \label{NU} \\
  \varepsilon^2 &\equiv& E^2/E_c^2 = I/I_c \simeq 1 \times 10^{-7} \;.
  \label{EPSILON}
\eea
We have listed the squared values as these typically arise as the
leading-order (LO) contributions the reasons being essentially
Lorentz and gauge invariance.

In view of the small parameters (\ref{NU}) and (\ref{EPSILON}) is
seems safe to assume that LO accuracy in $\nu^2$ and $\epsilon^2$
will suffice for the time being. Anticipating further increase in
laser power \cite{tajima:2002} we will, however, also consider
higher orders in this paper and confront the outcome numerically
with the exact (albeit implicit) one-loop results. This should
provide some intuition about the limitations of derivative and
weak-field expansions in cases where exact results are not
available and one has to rely entirely on the accuracy of the
expansions.

The paper is organised follows. We first go through a general
(mainly algebraic) discussion of the polarisation tensor where we
also give the integral representations. Section~\ref{sec:NLO}
discusses the next-to-leading (NLO) results while
Section~\ref{sec:LARGEO} extends these to order 80. Our findings
are then compared with an asymptotic analysis of the integral
representations in Section~\ref{sec:ANRES}. We finally present our
conclusions in Section~\ref{sec:CONCL}.

\section{General Analysis of $\Pi^{\mu\nu}$}

In order to evaluate (\ref{PI_SPEC1}) the following choice for the
basis vectors $\epsilon_i$ has been suggested in
\cite{bialynicka-birula:1970},
\bea
  b^\mu &\equiv& F^{\mu\nu}k_\nu \; , \\
  \bt^\mu &\equiv& \tilde{F}^{\mu\nu} k_\nu \; , \\
  c^\mu &\equiv& F^{\mu\nu} b_\nu \; ,
\eea
where, as usual, $\tilde{F}^{\mu\nu} = (1/2)
\epsilon^{\mu\nu\alpha\beta} F_{\alpha\beta}$ denotes the dual
field strength. The first two basis vectors are obviously
orthogonal to $k$,
\be
  b \cdot k = 0 = \bt \cdot k \; ,
\ee
while the last one is not,
\be
  c \cdot k = - b^2 \; .
\ee
However, if we follow \cite{ritus:1972} and define
\be
  d^\mu \equiv \left( g^{\mu\nu} - \frac{k^\mu k^\nu}{k^2} \right)
  c_\nu \equiv \mathbb{P}^{\mu\nu} c_\nu \equiv - \frac{1}{k^2}
  \epsilon^{\mu\nu\rho\sigma} k_\nu b_\rho \bt_\sigma \; ,
\ee
the vectors $b$, $\bt$ and $d$ constitute an orthogonal dreibein
which satisfies
\be
   \frac{d^\mu d^\nu}{d^2} + \frac{\bt^\mu \bt^\nu}{b^2} + \frac{b^\mu
   b^\nu}{b^2} = \mathbb{P}^{\mu\nu} \; .
\ee
The spectral decomposition (\ref{PI_SPEC1}) may then be rewritten
as
\be \label{PI_SPEC2}
  \Pi^{\mu\nu} = \Pi_0 \frac{d^\mu d^\nu}{d^2} + \Pi_+ \frac{\bt^\mu
  \bt^\nu}{b^2} + \Pi_- \frac{b^\mu b^\nu}{b^2} \; ,
\ee
and coincides with the one used in the monograph
\cite{dittrich:2000}. The eigenvalues $\Pi_i$ depend on all the
independent invariants one can form from $k^\mu$, the basis
vectors and the background field strength. For crossed fields,
however, most of these are either vanishing or dependent
quantities. In particular, we have
\bea
  \SCS &\equiv& - \frac{1}{4} F_{\mu\nu} F^{\mu\nu} = 0 \; , \label{S0} \\
  \SCP &\equiv& - \frac{1}{4} F_{\mu\nu} \tilde{F}^{\mu\nu} = 0 \label{PS0} \;
  .
\eea
We are thus left with only two basic invariants, $k^2$ and $b^2$,
such that the eigenvalues in (\ref{PI_SPEC2}) can be written as
\be \label{PII}
  \Pi_i \equiv k^2 P (k^2 , b^2) + \frac{b^2}{I_c} P_i (k^2 , b^2) \; , \quad
  i = 0, \pm \; ,
\ee
where $P$ and $P_i$ are dimensionless polynomials in $k^2$ and
$b^2$ (see below). Note that the $k^2$-term is the same for all
eigenvalues. According to (\ref{INDEX}) we have
\be
  k^2 = \omega^2 (1 - n^2) \; ,
\ee
which is negative in a modified vacuum ($n > 1$). To calculate
$b^2$ we note that the Maxwell energy-momentum tensor for the
crossed background fields is
\be
  T^{\mu\nu} = F^{\mu\lambda} {F_\lambda}^\nu - g^{\mu\nu} \SCS
  = F^{\mu\lambda} {F_\lambda}^\nu \; , \label{TMUNU}
\ee
whereupon $b^2$ simply becomes
\be
  b^2 (k) = - T^{\mu\nu} k_\mu k_\nu \; .
\ee
Using (\ref{INDEX}) again this translates into the expression
\be \label{B2H}
  b^2 = - \omega^2 [\mathcal{H} (1 + n^2) - 2n \vc{S} \cdot \vcb{k} - n^2
  \mathcal{H}_k] \; ,
\ee
where we have introduced the abbreviations
\bea
  \mathcal{H} &\equiv& T^{00} = \sfrac{1}{2} (E^2 + B^2) = F^2 \; , \label{T00} \\
  S^i &\equiv& T^{0i} = \epsilon^{ijk} E^j B^k \; , \label{POYNTING} \\
  \mathcal{H}_k &\equiv& (\vc{E} \cdot \vcb{k})^2 + (\vc{B} \cdot
  \vcb{k})^2 \; , \label{HK}
\eea
the first two of which represent the Maxwell energy-density and
the Poynting vector, respectively. Note that both these quantities
are directly related to the intensity,
\be
  \mathcal{H} = |\vc{S}| = I \; .
\ee
Expression (\ref{B2H}) for $b^2$ coincides with the quantity $z_k$
on p.\ 21 of \cite{dittrich:2000}. There is, however, an
alternative expression which nicely illustrates the kinematics
involved. Introducing the background 4-vector $K \equiv \Omega (1,
\vc{K})$, $\vc{K}^2 = 1$ and assuming background gauge $\partial
\cdot A = 0$ one finds
\be \label{B2KK}
  b^2 = b^2 (\theta) = - \omega^2 \, I \, (1 - n \cos \theta)^2 \le 0 \; ,
\ee
where $\theta$ is the angle between the probe and background
directions ($\vc{k}$ and $\vc{K}$, respectively). Note that $b^2$
becomes independent of $n$ for a perpendicular configuration
implying $b^2 (\pi/2) = - \omega^2 I$.

In order to obtain reasonably simple expressions for the
coefficient functions $P$ and $P_i$ multiplying $k^2$ and
$b^2/I_c$ in (\ref{PII}) we follow \cite{ritus:1972} and trade the
invariants $k^2$ and $b^2$ for the dimensionless parameters
\bea
  \lambda &\equiv& k^2 / m_e^2 = \nu^2 \, (1 - n^2) \le 0 \; , \label{LAMBDA} \\
  \kappa^2 &\equiv& - b^2/I_c m_e^2 = \epsilon^2 \nu^2 \, (1 - n \cos
  \theta)^2 \ge 0 \; . \label{KAPPA}
\eea
The eigenvalues of $\Pi^{\mu\nu}$ from (\ref{PII}) are then given
by the expressions
\bea
  \Pi_0 (\lambda, \kappa^2) &=& m_e^2 \, \lambda \, P(\lambda,
  \kappa^2) \; , \label{PI0} \\
  \Pi_\pm (\lambda, \kappa^2) &=& \Pi_0 (\lambda, \kappa^2) -
  m_e^2 \, \kappa^2 \, P_\pm (\lambda , \kappa^2) \label{PIPM} \;
  ,
\eea
implying $P_0 = 0$ in particular.  Using the integral
representations given in \cite{ritus:1972} the remaining
coefficient functions $P$ and $P_\pm$ may be compactly written as
\bea
  P (\lambda, \kappa^2) &=& P(\lambda, 0) - \frac{2\alpha}{\pi}
  \int_0^1 dx \, x \bar{x} \, f_1(z) \; , \label{P} \\
  P(\lambda , 0) &=& \frac{2\alpha}{\pi} \int_0^1 dx \, x \bar{x} \, \ln
  (1 - \lambda x \bar{x}) \; , \label{P0} \\
  P_\pm (\lambda , \kappa^2) &=& - \frac{\alpha}{3\pi}
  \, \kappa^{-4/3} \int_0^1 dx \, \frac{2 + (1 \pm 3) x \bar{x}}{(x
  \bar{x})^{1/3}} \, f^\prime (z) \label{PM} \; .
\eea
In the above, we have introduced the Feynman
parameters\footnote{The Feynman parameter $x$ is related to the
integration variables $\nu$ in \cite{dittrich:2000} and $v$ in
\cite{ritus:1972} via $\nu = 2x -1$ and $v = 1/x\bar{x}$,
respectively. In particular, $dv = (x - \bar{x})/(x\bar{x}^2)\,
dx$.} $x$ and $\bar{x} \equiv 1 - x$ and the variable
\cite{ritus:1972}
\be \label{Z}
  z = z(\lambda, \kappa, x) \equiv \frac{1 - \lambda x \bar{x}}{(\kappa x
  \bar{x})^{2/3}} \; ,
\ee
which is the argument of the auxiliary functions
\bea
  f(z) &\equiv& i \int_0^\infty dt \, e^{-izt -it^3/3} = \pi \,  [\Gi (z) + i
  \Ai(z)] \; , \label{F} \\
  f_1 (z) &\equiv& \int_0^\infty \frac{dt}{t} \, e^{-izt} \left(
  e^{-it^3/3} - 1   \right) \; . \label{F1}
\eea
$f^\prime$ denotes the derivative of $f$ with respect to $z$;
$\Gi$ and $\Ai$ are Scorer and Airy functions, respectively (see
e.g.\ \cite{abramowitz:64} or the Digital Library of Mathematical
Functions \cite{dlmf}).

As both $\Gi$ and $\Ai$ are real for real arguments we can
immediately infer from (\ref{PM}) and (\ref{F}) that the
eigenvalues $\Pi_\pm$ will develop imaginary parts determined by
the Airy function $\Ai(z)$ in the integrand. It is thus natural to
expect that also the index $n$ of refraction will become complex,
its imaginary part signalling absorption of photons by the vacuum.
Of course, the only physical interpretation of this phenomenon is
pair production.

%
%It is an intriguing question to ask whether the integrals in
%(\ref{P}--\ref{PM}) have an interpretation in terms of
%two-particle light-cone wave functions \cite{heinzl:2001} such
%that $x$ and $\bar{x}$ were longitudinal momentum fractions.

Enormous simplifications arise naturally for vanishing external
field, i.e.\ $\kappa^2 = 0$. In this case all eigenvalues become
degenerate, $\Pi_i \equiv \Pi = m_e^2 \, \lambda \, P$ and one
just obtains Schwinger's one-loop expression for the photon
self-energy \cite{schwinger:1951} in the form
\be
  \Pi^{\mu\nu} =  \Pi(k^2) \, \mathbb{P^{\mu\nu}}   = m_e^2 \,
  \lambda \, P(\lambda, 0) \, \mathbb{P^{\mu\nu}}   \; .
\ee
The scalar function $P(\lambda , 0)$ is given in (\ref{P0}) and
coincides with standard text book results  obtained via covariant
perturbation theory (see e.g.\ \cite{peskin:1995}, Ch.~7). For
small momenta it becomes
\be
  P(\lambda , 0) = - \frac{\alpha}{15\pi} \lambda + O(\lambda^2)
  \; ,
\ee
so that $\Pi(k^2) = O(k^4)$ for vanishing external field.

The next logical step is to expand the exponentials in (\ref{F})
and (\ref{F1}) in powers of $\lambda$ and $\kappa^2$. This amounts
to a derivative expansion in the probe field (keeping the
background fixed) and will be further pursued in the next section.

Before that let us conclude the general reasoning by determining
the dispersion relations for $k$. Adopting a plane wave ansatz for
the probe field, $f_{\mu\nu} = i(k_\mu \epsilon_\nu - k_\nu
\epsilon_\mu) \exp (i k \cdot x)$, and adding the Maxwell term
yields the wave equation
\be \label{WAVEEQ}
  \Box^{\mu\nu} \epsilon_\nu \equiv - (k^2 \mathbb{P}^{\mu\nu} - \Pi^{\mu\nu})
  \epsilon_\nu   = 0 \; ,
\ee
which has nontrivial solutions only if
\be \label{DET}
  \det \Box^{\mu\nu} (k)  = 0 \; ,
\ee
i.e.\ if an eigenvalue of $\Box^{\mu\nu}$ vanishes,
\be \label{DISP0}
  k^2 - \Pi_i (k^2 , b^2) \equiv h_i^{\mu\nu}(k^2 , b^2) k_\mu k_\nu = 0 \, .
\ee
Following \cite{liberati:2000} we have introduce effective metrics
$h_i^{\mu\nu}$ to make explicit that (\ref{DISP0}) represents
`modified light-cone conditions' \cite{dittrich:2000}. Inserting
(\ref{PII}) the three metrics become
\be
  h_i^{\mu\nu}(\lambda , \kappa^2) = g^{\mu\nu} \left[ 1 + P (\lambda, \kappa^2) \right] +
  T^{\mu\nu} P_i (\lambda, \kappa^2)/I_c \; .
\ee
As $P_0 = 0$ the metric $h_0^{\mu\nu}$ is conformally flat,
\be
  h_0^{\mu\nu}(\lambda , \kappa^2) = g^{\mu\nu} \left[ 1 + P (\lambda , \kappa^2) \right] \; ,
\ee
and does not modify the light-cone. Note in particular that in
contrast to a plasma of real charges weak crossed fields do not
generate a longitudinal photon so that $h_0^{\mu\nu}$ remains
unphysical.

The other two metrics, however, are physical and nontrivial
leading to a modified light propagation. As $h_1 \ne h_2$ this
implies `birefringence of the vacuum'. For vanishing external
fields ($\kappa^2 = 0$) all metrics merge into the conformal
metric $h_0^{\mu\nu} (\lambda , 0)$ which describes a standard
light-cone.

\section{NLO Derivative Expansion for the Probe}
\label{sec:NLO}

The integral representations (\ref{P}--\ref{PM}) for the
polarisation tensor may be expanded in powers of $\lambda$ and
$\kappa^2$ which are both $O(\nu^2)$. This derivative expansion
becomes particularly straightforward if we first expand in $\nu$
and only afterwards perform the integrals, a procedure adopted
throughout Sections \ref{sec:NLO} and \ref{sec:LARGEO}. Thus, we
first rewrite the derivative of (\ref{F}) and the second function
(\ref{F1}) in order to exhibit the dependence on $\kappa$ and
$\lambda$,
\bea
  f^\prime (z) &=& (\kappa x \bar{x})^{4/3} \int_0^\infty d\tau \,
  e^{i \lambda x \bar{x} \tau} e^{-i\tau - i \kappa^2 x^2 \bar{x}^2
  \tau^3/3} \; , \\
  f_1 (z) &=& \int_0^\infty \frac{d\tau}{\tau} e^{i \lambda x \bar{x}
  \tau} \, e^{-i\tau} \left( e^{- i \kappa^2 x^2 \bar{x}^2
  \tau^3/3} - 1 \right) \; .
\eea
Upon expanding the exponentials the integrals both over Feynman
parameters and proper time become elementary so that the
eigenvalues $\Pi_i$ to $O(\nu^4)$ are found to be
\bea
  \Pi_0/m_e^2 &=&  \phantom{- \frac{7 \alpha}{45 \pi} \kappa^2 \;\,}
  - \frac{\alpha}{15 \pi} \lambda^2  + \frac{\alpha}{105
  \pi} \lambda \kappa^2   \; , \\
  \Pi_+/m_e^2 &=&  - \frac{7 \alpha}{45 \pi} \kappa^2
  - \frac{\alpha}{15\pi} \lambda^2 - \frac{23 \alpha}{315 \pi}
  \lambda \kappa^2 - \frac{52 \alpha}{945 \pi} \kappa^4 \; ,  \\
  \Pi_-/m_e^2 &=& - \frac{4 \alpha}{45 \pi} \kappa^2  -
  \frac{\alpha}{15\pi} \lambda^2   - \frac{2 \alpha}{45 \pi} \; \, \lambda
  \kappa^2 - \frac{4 \alpha}{135 \pi} \kappa^4 \; .
\eea
Note that only $\Pi_\pm$ have LO contributions, $\Pi_\pm =
O(\kappa^2) = O(\nu^2)$, whereas $\Pi_0 = O(\lambda^2) =
O(\nu^4)$. To proceed we specialise to a head-on collision for
which birefringence becomes maximal, cf.\ (\ref{B2KK}) and
(\ref{KAPPA}). In this case, $\vc{k} \cdot \vc{K} = -1$ so that
$\kappa^2$ attains its maximum value,
\be \label{KAPPA_HO}
  \kappa^2 =  \epsilon^2 \nu^2 (1 + n)^2 \; .
\ee
The nontrivial dispersion relations $\Box_\pm \equiv k^2 - \Pi_\pm
= 0$ implicitly determine two nontrivial indices of refraction
$n_\pm$ as functions of the small parameters $\epsilon^2$ and
$\nu^2$. To solve the equations $\Box_\pm (n , \epsilon, \nu) = 0$
for the indices $n_\pm$ we write
\be
  n_\pm \equiv 1 + \Delta_\pm \equiv 1 + \frac{\alpha}{45 \pi} \,
  \epsilon^2 \, \delta_\pm \; .
\ee
The last expression takes into account that corrections to $n=1$
are due to nonvanishing external field intensity $\epsilon^2$ the
coupling to which is $O(\alpha)$. The remainders $\delta_\pm$ are
expected to be of order unity. For the value $\epsilon^2 =
10^{-7}$ from (\ref{EPSILON}) the prefactor is of order $10^{-11}$
so that at present the deviations from $n=1$ are extremely small.
It hence remains an experimental challenge to really measure them
\cite{heinzl:2006,zavattini:2005}.

If we expand the deviations according to
\be \label{DELEXP}
  \delta_\pm = \delta_{0\pm} + \delta_{2\pm} \, \nu^2 + O(\nu^4) \; .
\ee
we find at LO in $k^2$ or $\nu^2$ i.e.\ $O(\nu^0)$,
\bea
  \delta_{0+} &=& \frac{14}{1 - 7  \alpha \epsilon^2/45\pi} = 14
  \left\{ 1 + 7 \, \frac{\alpha \epsilon^2}{45 \pi}  + O(\alpha^2
  \epsilon^4) \right\} \; ,
  \label{DELTA0+} \\
  \delta_{0-} &=& \frac{8}{1 - 4  \alpha \epsilon^2/ 45\pi} =
   8 \left\{ 1 + 4 \, \frac{\alpha \epsilon^2}{45 \pi}  +
  O(\alpha^2 \epsilon^4) \right\}\; . \label{DELTA0-}
\eea
Note that the first terms on the right-hand side are exact to all
orders in the intensity $\epsilon^2$. An independent check of the
LO results based on the Heisenberg-Euler Lagrangian will be
performed in App.~A.

To the best of our knowledge, the  coefficients of order
$\epsilon^0$ in (\ref{DELTA0+}) and  (\ref{DELTA0-}) have first
been obtained in Toll's thesis \cite{toll:1952}\footnote{As we
have not been able to get hold of this unpublished work our
statement is based on the notes \cite{mcdonald:1986} where the
relevant figure of Toll's thesis is reproduced on p.33.} and
independently in \cite{baier:1967a,baier:1967b} (see also
\cite{brezin:1970b,narozhnyi:1968,adler:1971}). The NLO=$O(\nu^2)$
expressions are somewhat more complicated,
%\begin{widetext}
%
\bea
  \delta_{2+} &=& \left( \frac{416}{21} - \frac{184 \alpha}{45 \pi} +
  \frac{392 \alpha^2}{675 \pi^2} \right) \frac{\epsilon^2}{(1 -
  7\alpha\epsilon^2/45\pi)^4} \; , \label{DELTA2+} \\[10pt]
  \delta_{2-} &=& \left( \frac{32}{3} - \frac{64 \alpha}{45 \pi} +
  \frac{128 \alpha^2}{675 \pi^2} \right) \frac{\epsilon^2}{(1 -
  4\alpha\epsilon^2/45\pi)^4} \; . \label{DELTA2-}
\eea
%
%\end{widetext}
Again, these expressions are exact to all orders in $\epsilon^2$.
For later purposes it is useful to attach names to the different
factors,
\be \label{FACTOR}
  \delta_{2\pm} = c_{2\pm}(\alpha) \epsilon^2 s_{2\pm}
  (\alpha\epsilon^2) \; .
\ee
According to (\ref{DELTA2+}) and (\ref{DELTA2-}) the $c_{2\pm}$
are polynomials in $\alpha/45\pi$ while the $s_{2\pm}$ have the
series expansions
\bea
  s_{2+} &=& 1 + \frac{28\alpha}{45\pi} \epsilon^2 + \frac{98
  \alpha^2}{405 \pi^2} \epsilon^4 + O(\epsilon^6) \; , \label{S2+} \\
  s_{2-} &=& 1 + \frac{16\alpha}{45\pi} \epsilon^2 + \frac{32
  \alpha^2}{405 \pi^2} \epsilon^4 + O(\epsilon^6)
  \; . \label{S2-}
\eea
Note that both at LO and NLO the QED expansion parameter is
actually $\alpha/45\pi \simeq 5 \times 10^{-5}$ rather than
$\alpha$ itself. The coefficients $c_{2\pm}$ are dominated by the
leading terms of $O(\alpha^0)$ which both have a positive sign.
This implies that (at least to first nontrivial order in the
derivative expansion) the indices of refraction increase with
frequency so that we have normal (rather than anomalous)
dispersion.

Summarising the findings above we see the following pattern
emerging. Each of the two indices of refraction has a derivative
expansion in $\nu^2$,
\be \label{N_PM}
  n_\pm = 1 + \Delta_\pm = 1 + \frac{\alpha\epsilon^2}{45\pi} \sum_{l=0}^\infty \,
  \delta_{2l \pm} (\alpha, \epsilon) \; \nu^{2l} \; ,
\ee
where the $\delta_{2l}$ can be factorised by generalising
(\ref{FACTOR}),
\be \label{DELTA_2L}
  \delta_{2l\pm} \equiv c_{2l\pm} (\alpha) \; \epsilon^{2l}
  s_{2l\pm} (\alpha \epsilon^2/45\pi) \; .
\ee
Hence, the different coefficient functions have the generic
behaviour
\bea
  \delta_{2l} &=& O(\epsilon^{2l}) \; , \\
  c_{2l} &=& 1 + O(\alpha) \; , \\
  s_{2l} &=& 1 + O(\alpha\epsilon^2) \; ,
\eea
for both subscripts $\pm$. Keeping only the leading orders in
$\alpha$ in (\ref{DELTA_2L}), i.e.\ $\delta_{2l} \simeq c_{2l}(0)
\epsilon^{2l}$, results in a compact expression for the NLO
derivative expansion,
\be \label{DELTA_PM}
  \delta_\pm = 11 \pm 3 + \frac{320 \pm 96}{21}
  \epsilon^2 \nu^2 + O(\alpha) + O(\alpha \epsilon^2) \; .
\ee
Again, the relative plus sign between the first and second terms
signals normal dispersion. Extrapolating (\ref{DELTA_PM}) to all
orders adopting the same approximations we expect (\ref{DELEXP})
to become a function of $\epsilon\nu$ only,
\be \label{DELAPP}
  \delta \simeq \sum_{l=0}^\infty c_{2l}(0) (\epsilon \nu)^{2l} \;
  .
\ee
If we have a closer look at the coefficients $c_{0\pm}$ and
$c_{2\pm}$ (at $O(\alpha^0$)) we see that they actually increase,
the ratios being almost the same for both indices,
\be
  \frac{c_{2+}}{c_{0+}} \simeq \frac{416/21}{14} \simeq 1.4  \quad
  \mbox{and} \quad \frac{c_{2-}}{c_{0-}} \simeq \frac{32/3}{8} \simeq 1.3
  \; .
\ee
This is a first hint that our expansion in $\nu$ (or
$\epsilon\nu$) is only asymptotic which has to be expected upon
comparing with closely related derivative expansions of effective
actions \cite{dunne:2004,dunne:1999}. In the remainder of the
paper we will investigate this issue in detail.

%\section{(Semi-)Numerical Results}

\section{Large-Order Derivative Expansion for the Probe}
\label{sec:LARGEO}

\begin{figure}[!h]
\begin{center}
\includegraphics[scale=1]{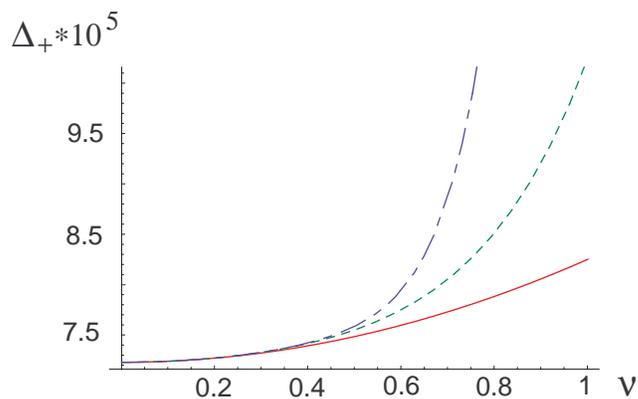}
\caption{\label{DELPLUS} Higher order results for the deviation
$\Delta_+$ of the index $n_+$  from unity as a function of $\nu =
\omega/m_e$ at fixed background intensity $\epsilon^2 = 0.1$. Full
line: second order; short-dashed line: sixth order; long-dashed
line: tenth order.}
\end{center}
\end{figure}

\begin{figure}[!h]
\begin{center}
\includegraphics[scale=1]{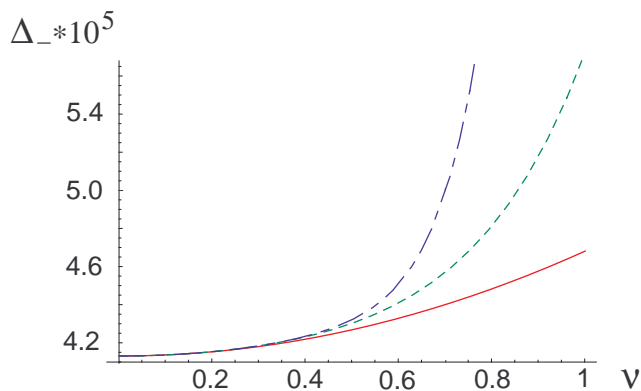}
\caption{\label{DELMINUS} Same as Fig.~\protect\ref{DELPLUS} for
$\Delta_-$.}
\end{center}
\end{figure}

Given the present day power of computer algebra systems we have
decided to actually check whether normal dispersion persists
beyond NLO in the derivative expansion. The answer is affirmative
as shown in the \texttt{Mathematica} plots of figures
\ref{DELPLUS} and \ref{DELMINUS}. They display the second, sixth
and tenth order in the derivative expansion of $\Delta = n - 1$ as
a function of $\nu$ at fixed background intensity $\epsilon^2 =
0.1$. The LO results (\ref{DELTA0+}) and (\ref{DELTA0-}) are given
by the common intercept of the different graphs. One clearly
identifies normal dispersion and a tendency for the curves to
diverge with increasing order.

As already stated the results of the preceding section indicate
strongly that the expansion (\ref{N_PM}) of the index of
refraction is asymptotic both in frequency and intensity. Hence,
for both the derivative and weak (external) field expansions we
expect \textit{factorial growth} of the expansion coefficients at
large orders.

To test this expectation numerically we expand $\Delta_\pm$ in
accordance with (\ref{N_PM}),
\be \label{DELTA_EXPAND}
  \Delta_\pm (\nu, \epsilon) = \frac{\alpha \epsilon^2}{45 \pi} \sum_{l \ge
  0} \delta_{2l\pm}(\epsilon) \, \nu^{2l} \; ,
\ee
with $\delta_{2l\pm}$ as given in (\ref{DELTA_2L}) and form the
ratio\footnote{We thank Paul Rakow for bringing this idea to our
attention \cite{rakow:2005}.}
\be \label{R_2L}
  r_{2l\pm} \equiv \frac{\delta_{2l+2,\pm}}{\epsilon^2 \,
  \delta_{2l\pm}} = \frac{c_{2l+2, \pm}(0)}{c_{2l,\pm}(0)} \left[
  1 + O(\alpha) + O(\alpha \epsilon^2)
  \right] \; .
\ee
The coefficients can (in principle) be determined by expanding the
integral representations (\ref{P}--\ref{PM}) using the auxiliary
functions (\ref{F}) and (\ref{F1}) (see next section). For simple
factorial growth, $\delta_{2l} \sim \Gamma (l)$, $r_{2l}$ would
depend linearly on $l$.

In what follows we want to ask the question what can be said about
the factorial growth \textit{without} using any knowledge of the
special functions (\ref{F}) and (\ref{F1}). That is, we perform a
brute-force derivative expansion before calculating any integrals,
trying to extend the method of the previous section to orders as
large as possible to gain a maximum of information. The philosophy
behind is our expectation that for realistic backgrounds such as
laser fields this type of derivative expansion will be one of the
main tools when analytical methods are not available. We thus
expect our method to provide results complementary to numerical
approaches like world-line Monte Carlo
\cite{gies:2001a,gies:2001b,gies:2005a,gies:2005b}.

In order to extract the asymptotic behaviour of $\delta_{2l}$ or
$r_{2l}$ with high accuracy it is obviously desirable to go to the
largest orders possible. Again, this is only feasible with the aid
of computer algebra and was performed with an optimised
\texttt{Mathematica} routine. Using a standard desktop PC we were
able to achieve a maximal order of $\ell=82$. The determination of
all 82/2 coefficients for a given value $\epsilon^2$ of intensity
takes about an hour.

\begin{figure}[h]
\begin{center}
\includegraphics[scale=1]{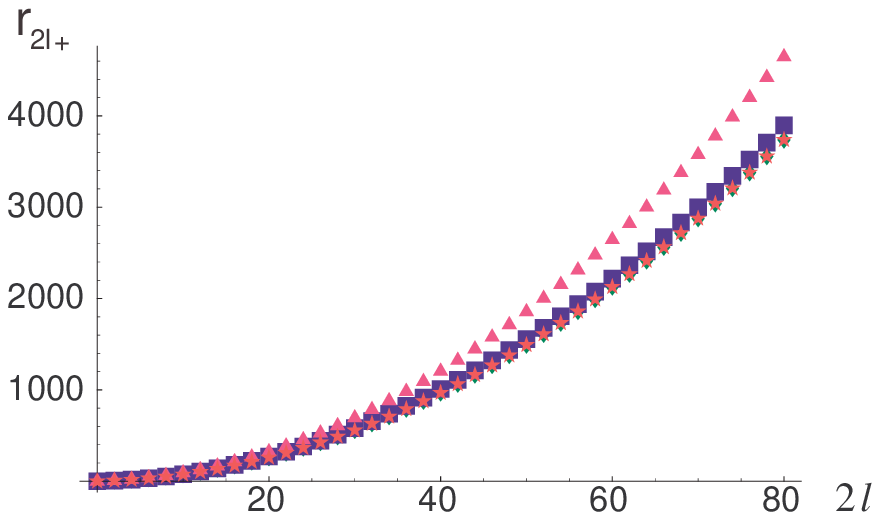}
\caption{\label{R2LPLUS} Successive coefficient ratio $r_{2l+}$ as
a function of order $2l$ for four different intensities,
$\epsilon^2 = 0.1$ ($\fulldiamond$), $\epsilon^2 = 1$
($\fullstar$), $\epsilon^2 = 100$ ($\fullsquare$) and $\epsilon^2
= 500$ ($\fulltriangle$).}
\end{center}
\end{figure}

\begin{figure}[h]
\begin{center}
\includegraphics[scale=1]{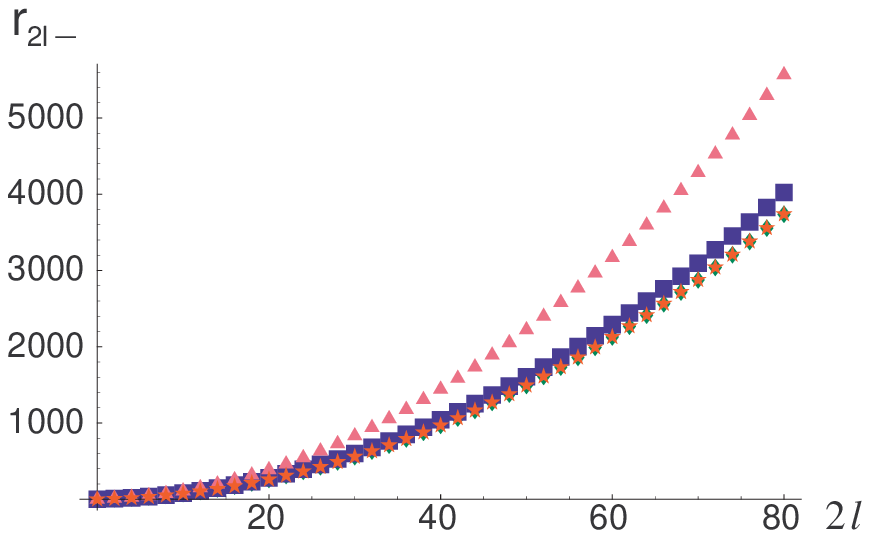}
\caption{\label{R2LMIN} Successive coefficient ratio $r_{2l-}$ as
a function of order $2l$ for four different intensities,
$\epsilon^2 = 0.1$ ($\fulldiamond$), $\epsilon^2 = 1$
($\fullstar$), $\epsilon^2 = 100$ ($\fullsquare$) and $\epsilon^2
= 500$ ($\fulltriangle$).}
\end{center}
\end{figure}

In Fig.s \ref{R2LPLUS} and \ref{R2LMIN} we have plotted the ratio
(\ref{R_2L}) as a function of order $2l$ for different values of
the intensity $\epsilon^2$ ranging from 0.1 (subcritical) to 500
(supercritical).

One notes the following: First, the $\epsilon$-dependence of
$r_{2l}$ is rather weak as expected from the discussion in the
preceding section and the second expression in (\ref{R_2L}). To
recognize corrections to the leading behavior which is independent
of $\epsilon$ one has to go to very large intensities and/or
orders. For instance, the graphs for $\epsilon^2 = 0.1$ and
$\epsilon^2 = 1$ basically are on top of each other. One needs
$\epsilon = O(10^2)$ to see 10\% deviations from the graphs with
small $\epsilon$ values. This is consistent with the coefficients
of $\epsilon^2$ in (\ref{R_2L}) being of order $10^{-3}$, cf.\
(\ref{S2+}) and (\ref{S2-}).

Second, and more importantly, the ratio $r_{2l}$ seems to depend
quadratically on $l$ which is simply explained by an asymptotic
behaviour $r_{2l} \sim \Gamma(2l)$. Let us try to perform a more
quantitative analysis. The following general ansatz
\be
\label{GAMMA_ASYMPT}
  c_{2l}(0) = \rho^{2l} \Big[ \Gamma(2l - \sigma) + \tau \, \Gamma(2l - \sigma -
  1)   \Big] \; ,
\ee
which is tailored after examples where exact results are available
\cite{dunne:1999,dunne:2002}, implies a ratio
\bea
  r_{2l} = \rho^2 \Big[ (2l)^2 + (1 - 2\sigma) \, 2l + \sigma^2 - \sigma - 2\tau
  \Big] \equiv a_2 \, (2l)^2 + a_1 \, 2l + a_0  \label{R_2L_FIT} \; .
\eea
Fitting the curves of Fig.s \ref{R2LPLUS} and \ref{R2LMIN} to
quadratic polynomials we find the $a_i$ values listed in
Table~\ref{R2LNUM}.
\begin{table}[!h]
\caption{\label{R2LNUM}Fitted values for the polynomial
coefficients in (\protect\ref{R_2L_FIT}).}
\begin{center}
\begin{tabular}{|l|llll|}
\hline
& $\epsilon^2 = 0.1$ & $\epsilon^2 = 1$ & $\epsilon^2 = 100$ & $\epsilon^2 = 500$\\
\hline
$a_{2+}$  & $0.56256(11)$ & $0.56295(35)$ & $0.607(11)$ & 0.8383(38) \\
$a_{1+}$  & $1.6892(29)$  & $1.6899(49)$  & $1.817(44)$ & 2.51(42) \\
$a_{0+}$  & $1.691(41)$   & $1.693(46)$   & $1.69(27)$  & 1.4(2.5) \\
\hline
$a_{2-}$  & $0.562535(27)$ & $0.56275(15)$ & $0.5865(15)$ & 0.699(11) \\
$a_{1-}$  & $1.68741(46)$  & $1.68795(16)$ & $1.7591(21)$ & 2.10(15) \\
$a_{0-}$  & $1.3926(36)$   & $1.3928(70)$  & $1.38(13)$   & 1.25(90)\\
\hline
\end{tabular}
\end{center}
\end{table}

The errors have been estimated by also fitting higher-order
polynomials and monitoring the change in the coefficients. For all
fits an extrapolation to $l \to \infty$ has been performed using
Laurent series techniques.

Via (\ref{R_2L_FIT}) the values of Table~\ref{R2LNUM} translate
into the asymptotic coefficients $\rho$, $\sigma$ and $\tau$
introduced in (\ref{GAMMA_ASYMPT}) as listed in Table~\ref{RST}.

\begin{table}[!h]
\caption{\label{RST}Numerical values for the polynomial
coefficients in (\protect\ref{GAMMA_ASYMPT}).}
\begin{center}
\begin{tabular}{|l|llll|}
\hline
& $\epsilon^2 = 0.1$ & $\epsilon^2 = 1$ & $\epsilon^2 = 100$ & $\epsilon^2 = 500$\\
\hline
$\rho_+$    & $0.75004(15)$ & $0.75030(47)$  & $0.779(14)$   & $0.9156(42)$ \\
$\sigma_+$  & $-1.0014(23)$ & $-1.009(34)$   & $-0.9967(91)$ & $-1.00(24)$ \\
$\tau_+$    & $-0.501(65)$  & $-0.502(69)$   & $-0.40(35)$   & $0.2(2.2)$ \\
\hline
$\rho_-$    & $0.750023(36)$  & $0.75017(20)$  & $0.7658(20)$  & $0.836(13)$ \\
$\sigma_-$  & $-0.99983(34)$  & $-0.99973(26)$ & $-0.9997(20)$ & $-1.002(84)$ \\
$\tau_-$    & $-0.2380(52)$   & $-0.238(12)$   & $-0.18(22)$   & $0.11(99)$ \\
\hline
\end{tabular}
\end{center}
\end{table}

One notes that $\rho$ and $\sigma$ are the same for both indices
of reflection i.e.\ $\rho_+ = \rho_- \equiv \rho$ and $\sigma_+ =
\sigma_- \equiv \sigma$ while $\tau_+ \ne \tau_-$. Furthermore,
$\sigma$ seems to be independent of intensity (unlike $\rho$ and
$\tau$). We can even guess the following `analytic' values for
$\rho$ and $\sigma$,
\be \label{RHOSIGMA}
  \rho = 3/4 \; , \quad \mbox{and} \quad \sigma = -1 \; .
\ee
For $\epsilon^2 \lesssim 1$ these values for $\rho$ and $\sigma$
should be quite accurate. Not unexpectedly, the largest errors
reside in the subleading coefficients $\tau_\pm$. In view of this
we refrain from any further guesses and just quote the
$\tau$-values of Table~\ref{RST}.

With the coefficients being determined let us plug the ansatz
(\ref{GAMMA_ASYMPT}) into (\ref{DELTA_EXPAND}) adopting the
approximation (\ref{DELTA_2L}) so that the expansion parameter
becomes $\epsilon \nu$ rather than $\nu$. This should be fine for
$\epsilon^2 \lesssim 1$. As our ansatz (\ref{GAMMA_ASYMPT}) is
supposed to hold only asymptotically for large orders we cannot
expect to describe the low orders accurately. We nevertheless
proceed by eliminating the Gamma functions using the integral
expression
\be
  \rho^{z} \, \Gamma(z) = \int_0^\infty ds \, \frac{e^{-1/\rho
  s}}{s^{z+1}} \; .
\ee
This transforms the required summations over $l$ into geometric
series and yields the following integral representation for
(\ref{DELTA_EXPAND}),
\be \label{DELINTREP}
  \Delta_\pm (\epsilon, \nu) = N_\pm \, \frac{\alpha \epsilon^2}{45\pi} \int_0^\infty ds \,
  \frac{s}{s^2 - (\epsilon\nu)^2} \left[ (\rho s)^\sigma + \tau_\pm
  (\rho s)^{\sigma + 1} \right] \, e^{-1/\rho s} \; ,
\ee
where we have allowed for an undetermined scale $N_\pm$. As it
stands the integral is ambiguous as there are poles on the real
axis. The left-hand side may be viewed as originating from the
derivative (or $\epsilon\nu$) expansion which has real
coefficients. It is therefore tempting to interpret
(\ref{DELINTREP}) as the following dispersion integral (see e.g.\
\cite{bjorken:1965}, Ch.~18),
\be \label{DISPREL}
  \mbox{Re} \, \Delta_\pm (\epsilon\nu) = N_\pm \, \frac{2}{\pi} \;
  \mathfrak{P} \int_0^\infty ds \, \frac{s}{s^2 -
  (\epsilon\nu)^2} \; \mbox{Im} \, \Delta_\pm (s) \; ,
\ee
where the pole ambiguity has been resolved in terms of the Cauchy
principal value denoted by $\mathfrak{P}$. The right-hand side
contains a `nonperturbative imaginary part' \cite{dunne:2002},
\be
  \Im \, \Delta_\pm (\epsilon\nu) = N_\pm \frac{\alpha \epsilon^2}{90}
  \, (\epsilon \nu \rho)^\sigma \, (1 + \tau_\pm \epsilon \nu \rho)
  \, e^{-1/\epsilon \nu \rho} \; ,
\ee
i.e.\ an exponential that cannot be obtained from a perturbative
expansion in $\epsilon\nu$. Inserting the universal values
(\ref{RHOSIGMA}) and $\tau_\pm$ from Table~\ref{RST} this becomes
\be \label{IMD_NUM}
  \Im \, \Delta_\pm (\epsilon \nu) = N_\pm \frac{\alpha \epsilon^2}{90}
  \, \frac{4}{3\epsilon \nu} \, \left( 1 - \frac{3 \epsilon \nu}{4}
  \left\{ 0.50 \atop 0.24 \right\}  \right) \, e^{-4/3\epsilon \nu}
  \; .
\ee
Note that the small parameter involved seems to be
$3\epsilon\nu/4$.

Summarising we can say that our large-order derivative expansion
has provided us with a quantitative formula for the asymptotics of
the series which by means of (\ref{DISPREL}) could be turned into
a statement about the nonperturbative imaginary part. As already
stated, this should be directly related to absorption and pair
production. In what follows we will check our findings by an
analytic discussion of the integral representations of the
eigenvalues $\Pi_i$, in particular of (\ref{PM}).

\section{Analytic Results}
\label{sec:ANRES}

Proceeding analytically is equivalent to analysing the (derivative
of the) auxiliary function $f(z)$ introduced in (\ref{F}). The
leading orders for both weak and strong external fields have
already been investigated by Narozhnyi \cite{narozhnyi:1968}. In
this section we want to go further and discuss arbitrary large
orders in the low-frequency, weak-field expansion of the
eigenvalues $\Pi_i$. Nevertheless, to keep things manageable we
follow Narozhnyi's example and adopt an approximation that leads
to substantial simplifications. Namely, we neglect vacuum
modifications \textit{in expressions involving the eigenvalues}
$\Pi_i (\lambda, \kappa)$ by setting $n=1$ or, equivalently,
$\lambda = 0$ in their arguments. From (\ref{PI0}) and
(\ref{PIPM}) we thus have the approximate identities
\bea
  \Pi_0 (\lambda, \kappa^2) &\simeq& \Pi_0(0, \kappa^2) = 0 \; , \\
  \Pi_\pm (\lambda, \kappa^2) &\simeq& \Pi_\pm (0, \kappa^2) = - m_e^2
  \kappa^2 P_\pm (0, \kappa^2) \; .
\eea
According to (\ref{KAPPA_HO}) we have in the same approximation,
\be \label{KAPPA_APP}
  \kappa \simeq 2\epsilon\nu \; ,
\ee
which turns our eigenvalues into functions of the product
$\epsilon\nu$ in line with our discussion of the previous section.
The upshot of all this is that the dispersion relations
(\ref{DISP0}) simplify drastically so that one is left with the
task to determine $P_\pm$ in
\be \label{DISPAPP}
  \Delta_\pm (\kappa, \lambda) \simeq \Delta_\pm (\epsilon\nu)
  \simeq 2 \epsilon^2 P_\pm (\epsilon \nu) \; .
\ee
As an aside we remark that according to (\ref{B2KK}),
(\ref{LAMBDA}) and (\ref{KAPPA}) Narozhnyi's approximation is
expected to be particularly good if probe and background are
perpendicular as $b^2$ and hence $\kappa$ are indeed independent
of $n$ in this case\footnote{with the replacement $2\epsilon \nu
\to \epsilon \nu$ understood in (\ref{KAPPA_APP})}. The
perpendicular setup has recently been suggested as a means to look
for `vacuum diffraction' \cite{dipiazza:2006}.

To evaluate (\ref{DISPAPP}) we want to calculate $P_\pm$ as a
power series in $\epsilon\nu$ using the integral representation
(\ref{PM}). From the definition of $\kappa^2$ in (\ref{KAPPA}) or
(\ref{KAPPA_APP}) and  of $z(\lambda, \kappa, x \bar{x})$ in
(\ref{Z}) it is clear that an expansion in $\epsilon\nu$
corresponds to determining the large-$z$ asymptotics of
\be \label{FPRIME1}
  f' (z) = \pi \, \Gi' (z) + i \pi \, \Ai' (z) \; ,
\ee
which appears in the integrand of (\ref{PM}). Using asymptotic
expansions for $\Ai(z)$ given in \cite{abramowitz:64} and $\Gi(z)$
in \cite{gil:2002} one obtains for (\ref{FPRIME1})
\be \label{FPRIME2}
  f' (z) = - \frac{1}{z^2} - \frac{1}{z^5} \sum_{k=0}^\infty
  \frac{g_k}{z^{3k}} - \frac{i}{2} \, \pi^{1/2} z^{1/4} e^{-\zeta}
  \sum_{k=0}^\infty (-1)^k A_k \, \zeta^{-k} \; ,
\ee
with the abbreviations
\bea
  \zeta &\equiv& \frac{2}{3} z^{3/2} \; , \\
  g_k &\equiv& \frac{(3k+4) \, \Gamma(3k+3)}{3^k \, \Gamma(k+1)} \; , \\
  A_k &\equiv& - \frac{6k+1}{6k-1} \; \frac{\Gamma(3k + 1/2)}{54^k \label{AK} \,
  k! \, \Gamma(k+1/2)} \; .
\eea
The expression (\ref{FPRIME2}) now has to be plugged in the
integral (\ref{PM}). Throughout the subsequent calculations we use
Narozhnyi's approximation which implies
\be
  z \simeq (\kappa x \bar{x})^{-2/3} \label{Z_APP} \; , \quad \zeta
  \simeq 2/3\kappa x\bar{x} \; .
\ee
In what follows we will separately calculate real and imaginary
parts of $P_\pm$.

\subsection{Calculation of the Real Part}

Using (\ref{KAPPA_APP}), (\ref{Z_APP}) and the integral
representation of the Beta function,
\be
  \int_0^1 dx \, x^{n-1} \, \bar{x}^{m-1} = B(n , m) =
  \frac{\Gamma(n)\, \Gamma(m)}{\Gamma(n+m)} \; ,
\ee
the real part of $P_\pm$ can be obtained in closed form,
\be \label{REPM}
  \fl \Re \, P_\pm (\kappa) =  \frac{\alpha m^2}{3 \pi}  \sum_{l \ge 0}
  G_l \, \kappa^{2l} \Big\{ 2 B(2l+2 , 2l+2) + (1 \pm 3) \, B(2l+3, 2l+3)
  \Big\} \; .
\ee
Here, we have introduced the new expansion coefficients
\be
  G_l \equiv \left\{ \begin{array}{ll}
                     1 \; ,  &  l = 0 \\
                     g_{l-1} \; ,  & l > 0
                     \end{array}
  \right.
\ee
The leading term ($l=0$) in the expansion (\ref{REPM}) yields for
(\ref{DISPAPP})
\be
  \Delta_\pm (\epsilon\nu) = 2 \epsilon^2 P_\pm (\epsilon\nu) \simeq (11 \pm 3)
  \frac{\alpha\epsilon^2}{45\pi}  \; ,
\ee
which reassuringly is consistent with (\ref{DELTA_PM}).

\begin{figure}[!h]
\begin{center}
\includegraphics[scale=0.5]{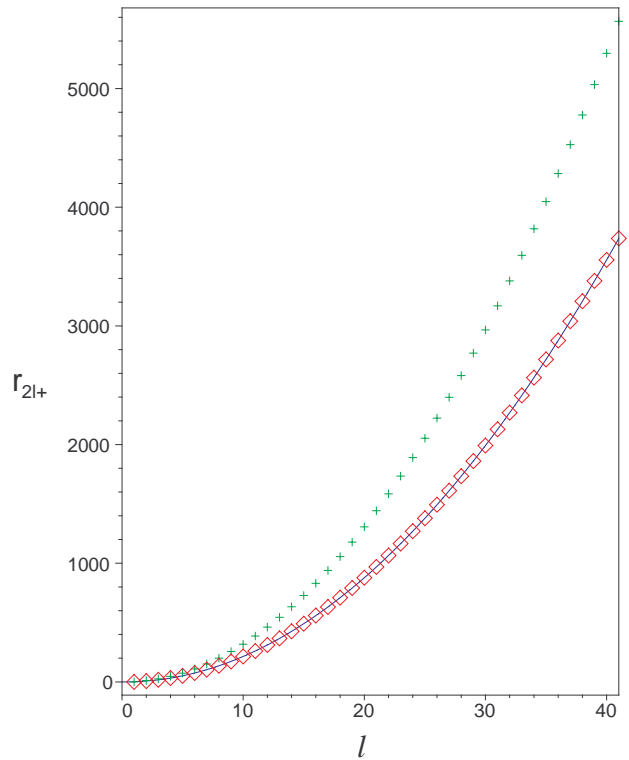}
\caption{\label{R2LFIT} Coefficient ratio $r_{2l+}$ as a function
of order $l$. Diamonds: numerical values for $\epsilon^2 = 0.1$;
crosses: numerical values for $\epsilon^2 = 500$ (both as in
Fig.~\protect\ref{R2LPLUS}); full line: analytical ratio from
(\protect\ref{R2LANAL}).}
\end{center}
\end{figure}

It is the presence of the Beta functions in (\ref{REPM}) that
causes factorial growth. To analyse this in the spirit of the
previous section we expand the corrections to the indices of
refraction according to (\ref{N_PM}) and (\ref{DELAPP}),
\be
  \Re \, \Delta_\pm \equiv 2 \epsilon^2 \Re \, P_\pm =
  \frac{\alpha\epsilon^2}{45\pi} \sum_{l \ge 0} c_{2l,\pm}(0)
  (\epsilon\nu)^{2l} \; .
\ee
Comparing with (\ref{REPM}) we find the coefficient ratios
\be \label{R2LANAL}
 r_{2l\pm} = 4 \, \frac{G_{l+1}}{G_l} \frac{2 B(2l+4, 2l+4) + (1 \pm 3) B(2l+5,
 2l+5)}{2 B(2l+2, 2l+2) + (1 \pm 3) B(2l+3, 2l+3)} \; .
\ee
These can be evaluated exactly to give
\bea
   r_{2l+} &=& \frac{2 (6l+13) (3l+4) (3l+2) (2l+3)
  (l+1)}{(6l+7) (4l+9) (4l+7)} \nn \\[5pt]
  &=& \frac{9}{16} \,  (2l)^2 + \frac{54}{32} \, 2l + \frac{452}{256} -
  \frac{372}{256} \frac{1}{2l} + O(1/l^2) \; , \label{R2LANALPLUS} \\[10pt]
  r_{2l-} &=& \frac{(6l+14) (3l+2) (2l+3)
  (l+1)}{(4l+9) (4l+7)} \nn \\[5pt]
  &=& \frac{9}{16} \, (2l)^2 + \frac{54}{32} \, 2l + \frac{356}{256} -
  \frac{12}{256} \frac{1}{2l} + O(1/l^2) \; . \label{R2LANALMIN}
\eea
Obviously, as the $r_{2l\pm}$ are independent of intensity they
cannot describe all curves of Fig.s \ref{R2LPLUS} and \ref{R2LMIN}
which, after all, \textit{are} intensity dependent. However, we
expect good agreement between numerical and analytical values of
$r_{2l\pm}$ for small intensities $\epsilon^2 \lesssim 1$. This is
nicely corroborated by the graphs of Fig.~\ref{R2LFIT} where the
results for $\epsilon^2 = 0.1$ are on top of each other for all
$l$. This near-perfect agreement is due to the fact that upon
neglecting the $\epsilon$ dependence of the coefficients
(\ref{DELTA_2L}) (and within Narozhnyi's approximation)
\textit{all} expansion coefficients are basically factorials, cf.\
(\ref{REPM}).

Matching (\ref{R2LANALPLUS}) and (\ref{R2LANALMIN}) with
(\ref{R_2L_FIT}) we read off the coefficients
\bea
  a_2 &=& \frac{9}{16} = 0.5625 \; , \\
  a_1 &=& \frac{54}{32} = 1.6875 \; , \\
  a_{0+} &=& \frac{452}{256} = 1.765625 \; , \\
  a_{0-} &=& \frac{356}{256} = 1.390625 \; ,
\eea
which compare favourably with the entries of Table~\ref{R2LNUM}
for $\epsilon^2 \lesssim 1$. The $a_i$ translate exactly into the
asymptotic parameters $\rho = 3/4$ and $\sigma = -1$ we had
guessed in (\ref{RHOSIGMA}) while the $\tau$-parameters become
\bea
  \tau_+ &=& - \frac{41}{72} = -0.5694 \; , \\
  \tau_- &=& - \frac{17}{72} = -0.2361 \; .
\eea
Again, these agree satisfactorily with the entries of
Table~\ref{RST} for small $\epsilon$, in particular for subscript
`minus'.

It remains to rewrite our nonperturbative imaginary part
(\ref{IMD_NUM}) by replacing the numerical $\tau_\pm$ by their
analytic counterparts just obtained,
\be \label{IMDEL}
  \Im \, \Delta_\pm (\epsilon \nu) = N_\pm
  \frac{\alpha\epsilon^2}{90} \, \frac{4}{3\epsilon \nu} \, \left(
  1 - \frac{3 \epsilon \nu}{4} \frac{29 \pm 12}{72}
  \right) \, e^{-4/3\epsilon \nu} \; .
\ee
Let us finally try to check this by a direct calculation of the
imaginary part.

\subsection{Calculation of the Imaginary Part}

The determination of the imaginary part of $P_\pm$ is more
involved as the integrand (\ref{PM}) contains an exponential that
cannot be expanded in powers of $x\bar{x}$. Explicitly, the
integrand is of the form
\be
  e^{-\zeta} \, (x \bar{x})^n \simeq \exp(-2/3\kappa x \bar{x}) \,
  (x \bar{x})^n \; ,
\ee
which is to be integrated from 0 to 1. In order to proceed
analytically we note that the product $x\bar{x}$ is peaked at the
value 1/4. Thus, it seems feasible to perform the integral via
saddle point approximation which results in the formula
\be \label{SADDLE}
  I(a,b) \equiv \int_0^1 dx \, e^{-ax \bar{x}} (x\bar{x})^b \simeq
  \left( \frac{\pi}{16a + 4b} \right)^{1/2} 4^{-b} \, e^{-4a} \; .
\ee
Here, the first prefactor has been obtained by extending the
integration to the whole real axis in order to have a Gaussian
integral. In our case, the parameters $a$ and $b$ are given by
\be
  a = 2/3\kappa \simeq 1/3\epsilon\nu \quad \mbox{and} \quad b = k \pm 1/2 \; , \; k =
  0, 1, 2, \ldots \; .
\ee
Comparison with a numerical evaluation for $k=0 \ldots 5$ shows
that the error of the approximation is about 10\% for $a=1$, 1\%
for $a=5$ and $0.1\%$ for $a=10$. Hence, for large $a$ (which we
have) the formula (\ref{SADDLE}) should work very well.

Having gained sufficient confidence in our saddle point
approximation we rewrite the imaginary part of (\ref{PM}) using
(\ref{FPRIME2}),
\be
  \fl \Im \, P_\pm \simeq  \frac{\alpha}{6\pi^{1/2}} \, \kappa^{-3/2}
  \sum_{k\ge0} (-1)^k \, A_k \, \Big\{ 2 \, I(a, k-1/2) + (1 \pm 3) \,
  I(a, k+1/2) \Big\} \; ,
\ee
with $A_k$ as defined in (\ref{AK}). Employing our integration
formula (\ref{SADDLE}) we find
\be \label{BSERIES}
  \Im \, P_\pm \simeq  \frac{\alpha}{24} \, \sqrt{\frac{3}{2}} \,
  \frac{1}{\kappa} \, e^{-8/3\kappa} \, \sum_{k\ge 0} B_{k\pm} \left(
  -\frac{3\kappa}{8}\right)^k  \; ,
\ee
where the new expansion coefficients are given by the somewhat
lengthy expression
\be
  B_{k\pm} \equiv A_k \, \left\{ \frac{9 \pm 3}{4} +
  \frac{72k}{(6k-7)(6k+1)} \left[ \frac{9\pm3}{4} (k-1) +
  \frac{-7\pm3}{8} \right] \right\} \; .
\ee
Admittedly, this is not too illuminating so we evaluate the series
(\ref{BSERIES}) to order $\kappa^2$ or $(\epsilon\nu)^2$ and use
(\ref{DISPAPP}) to determine
\be \label{IMDELEX}
  \fl \Im \, \Delta_\pm \simeq N_{\pm,\mathrm{sp}} \,
  \frac{\alpha\epsilon^2}{90} \, \frac{4}{3\epsilon\nu} \, e^{-4/3\epsilon\nu} \,
  \left\{ 1 + \frac{25\mp 12}{72} \frac{3\epsilon\nu}{4} +
  \frac{265 \mp 168}{2 \times 72^2} \left( \frac{3\epsilon\nu}{4} \right)^2
  \ldots \right\}  .
\ee
In the above we have introduced the saddle point normalisation
factor\footnote{The leading term in (\ref{IMDELEX}) has already
been determined by Ritus \cite{ritus:1972}. His normalisation
differs from ours by a factor of 2, a discrepancy we have not been
able to trace. The most plausible explanation seem to be the
ambiguities in determining the prefactor in the saddle point
approximation.}
\be
  N_{\pm,\mathrm{sp}} \equiv (3 \pm 1) \frac{45}{32}
  \sqrt{\frac{3}{2}} \; .
\ee
We refrain from identifying this with the analogous factor in
(\ref{IMDEL}) as the approximation schemes involved do not allow
for a direct comparison of the normalisation. The latter will be
fixed in a moment by means of the dispersion relation
(\ref{DISPREL}).

Putting these niceties aside we stress that we find perfect
agreement between (\ref{IMDEL}) and (\ref{IMDELEX}) concerning the
LO dependence on powers of $\epsilon\nu$ and the nonperturbative
exponential. The only discrepancy resides in the
$\tau$-coefficients multiplying the sub\-leading terms of order
$\epsilon\nu$. In view of the fairly different approximations
employed this is probably not too surprising.

\section{Discussion and Conclusion}
\label{sec:CONCL}

\begin{figure}
\begin{center}
\includegraphics[scale=0.6]{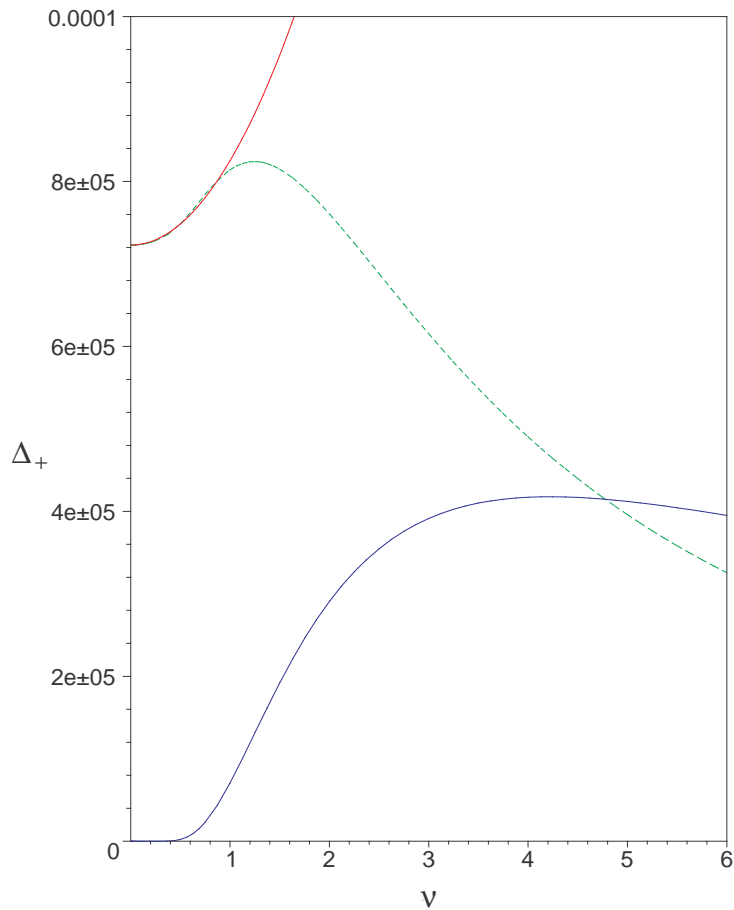}
\caption{\label{REIMDELTA} Real and imaginary part of the
deviation $\Delta_+$ of the index $n_+$ from unity (dashed and
lower full curve, respectively). The upper full curve bending
upwards is the series expansion of $\Delta_+$ to second order
(same curve as in Fig.\protect\ref{DELPLUS}).}
\end{center}
\end{figure}

With the consequences of our large-order analysis confirmed
analytically we can go even one step further.  We can use the
Kramers-Kronig relation (\ref{DISPREL}) to actually
\textit{define} the real part of the indices of refraction
nonperturbatively, i.e.\ without relying on the derivative
expansion. To this end we take the leading order of the imaginary
part (\ref{IMDEL}) or (\ref{IMDELEX}) as a \textit{model} by
writing
\be
  \Im \, \Delta_\pm (\epsilon, \nu) = N_\pm \,
  \frac{\alpha\epsilon^2}{90} \, \frac{4}{3\epsilon \nu} \,
  e^{-4/3\epsilon \nu} \; .
\ee
Plugging this into the the dispersion relation (\ref{DISPREL}) we
can actually perform the principal value integral analytically with
the result
\be \label{EI}
  \Re \, \Delta_\pm = N_\pm \, \frac{\alpha\epsilon^2}{90\pi} \,
  \frac{4}{3\epsilon \nu} \Big\{ \Ei (4/3\epsilon\nu) \,
  e^{-4/3\epsilon \nu} - \Ei (-4/3\epsilon\nu) \, e^{4/3\epsilon \nu}
  \Big\} \; .
\ee
Matching the perturbative small-$\epsilon\nu$ behaviour fixes the
normalisation to be
\be
  N_\pm = 11 \pm 3 \; .
\ee
Interestingly, in (\ref{EI}) exponential integrals Ei appear
multiplied by exponentials, constituting a paradigm example of
functions displaying factorial growth expansion coefficients. With
the real part thus determined we summarise our findings in
Fig.~\ref{REIMDELTA} which shows both real and imaginary part of
$\Delta_+$ as a function of frequency $\nu$ for fixed $\epsilon^2
= 0.1$. We have added the series expansion to second order (the
full line of Fig.\ref{DELPLUS}) which coincides well with the
exact real part for small $\nu$ where the factorial growth is not
visible yet.

The graph for $\Delta_-$ looks almost identical with a slight
shift in the vertical scale due to the difference in
normalisation. Hence, we refrain from producing an extra plot.

By construction, the real and imaginary parts of $\Delta$ (thus also
of $n = 1 + \Delta$) are related by the Kramers-Kronig relation
(\ref{DISPREL}) with $\Im \, n \ne 0$ signalling absorption, i.e.\
pair production. By looking at Fig.~\ref{REIMDELTA} we see that the
latter sets in roughly at the critical value of $\nu = 1$. Here,
$\Re \, n$ attains its maximum and decreases for $\nu \gtrsim 1$
while $\Im \, n$ increases. Mathematically, we may state
\be
  \begin{array}{lll}
  \displaystyle \pad{\Re \, n_\pm}{\omega} > 0  & \mbox{for} \; \omega \ll
  m_e \; , & \quad \mbox{(normal dispersion)} \\[10pt]
  \displaystyle \pad{\Re\, n_\pm}{\omega} < 0 & \mbox{for} \; \omega
  \gtrsim m_e \; , & \quad \mbox{(anomalous dispersion)}
  \end{array}
\ee
where we have reinstated physical units (recall that $\omega$ is
the probe frequency). We conclude that it is a consequence of the
Kramers-Kronig relations based on the fundamental principle of
causality that absorption is intimately connected to anomalous
dispersion.

The physics involved can be wrapped up as follows. We have
analysed the influence of crossed background fields on the
propagation of light (e.g.\ laser beams). Using exact integral
representations for the eigenvalues of the polarisation tensor we
found that to all orders in probe frequency and background
intensity there is birefringence of the vacuum induced by the
crossed background fields. The effect can be described in terms of
background dependent effective metrics $h_\pm^{\mu\nu}$ implying
dispersion relations $h_\pm^{\mu\nu} k_\mu k_\nu = 0$ which
describe distorted light-cones. Solving for the indices of
refraction $n_\pm$ one finds that they are frequency dependent,
starting out with normal dispersion. At critical energy and
intensity anomalous dispersion sets in together with absorption
due to pair production.

At present an experiment is being designed that plans to measure
vacuum birefringence using a high power laser background probed by
x-ray beams \cite{heinzl:2006}. The experiment is quite demanding
as one has to measure ellipticity signals with a sensitivity at
the order of $10^{-11}$ for presently available probes and lasers.
As the signal is proportional to $\nu^2 \epsilon^4$ it may be
readily enhanced by increasing probe frequency and background
intensity. Within the next few years one expects a reduction of
the required sensitivity down to $10^{-4}$. Normal dispersion is
an NLO effect, hence implies a signal of order $\nu^4 \epsilon^4$
requiring a sensitivity of $10^{-8}$ within the envisaged
scenario. This is still within the theoretical limits of
measurability \cite{alp:2000}. Pair production shows up in terms
of a nonvanishing imaginary part or as anomalous dispersion of the
real part. To become observable both effects require parameters
$\nu$ and $\epsilon$ close to their critical values. For instance,
if $\epsilon \simeq 10^{-2}$ then $\Im \Delta \simeq 10^{-11}$ for
$\nu \simeq 0.8$. Even these moderate values cannot be attained at
present so that it is presumably more reasonable to look for
positrons rather than optical signals to detect pair production.

From a theorist's point of view one should also consider two-loop
corrections and the influence of nonconstant backgrounds. The
two-loop corrections to the Heisenberg-Euler Lagrangian (cf.\
App.~A) have been calculated by Ritus \cite{ritus:1976}. They
amount to a replacement of the LO coefficients of (\ref{DELTA_PM})
according to
\be
  \left\{ 14 \atop 8 \right\} \to \left\{ {14 \; (1 + 1315\alpha/252
  \pi)}   \atop {8 \; (1 + 40\alpha / 9 \pi)} \right\} \; .
\ee
These are both one-percent corrections to the LO=$O(\nu^0)$
behaviour in (\ref{DELTA_PM}).

Regarding nonconstant backgrounds it is possible to slightly relax
the crossed-field assumption in a perfectly controlled manner.
Laser beams may be more realistically described as
\textit{Gaussian beams} rather than plane waves. The former have a
Gaussian profile in transverse direction of `waist size' $w_0$ and
a Lorenz profile in the longitudinal direction characterized by
the `Rayleigh length' $z_0$. One can form the small dimensionless
parameter $\triangle \equiv w_0/2z_0 \ll 1$ \cite{bulanov:2004}
which describes the deviation from the crossed-field limit
corresponding to $\triangle = 0$. Naturally, one expects that
there will be $O(\triangle)$-corrections to the results presented
here.

It is this context of nonconstant backgrounds where the intuition
gained in this paper is expected to pay off. In this more realistic
case, we cannot hope to have any exact analytical results available.
Thus it is important to know both the region of validity and the
limitations of derivative and weak-field expansions. From our
results they are both expected to break down if the product of
frequency squared and intensity becomes $\omega^2 I = O(m_e^6/e^2)$,
or, in dimensionless units, $\epsilon^2 \nu^2 = O(1)$ (see
Fig.~\ref{REIMDELTA}). With the present values of $\epsilon^2 \nu^2
\simeq 10^{-13}$ and those expected in the near future ($\epsilon^2
\nu^2 \simeq 10^{-10}$), however, one is definitely on the safe side
where (asymptotic) expansion methods make perfect sense.

\newpage

\ack It ia a pleasure to thank Arsen Khvedelidze, Martin Lavelle,
David McMullan, Paul Rakow, Roland Sauerbrey, Igor Shovkovy and
Andreas Wipf for very useful discussions. This research was
carried out while OS was a PPARC postdoctoral research fellow at
the University of Plymouth.

\appendix

\section{Heisenberg-Euler Analysis}

In this appendix we will check our LO results (\ref{DELTA0+}) and
(\ref{DELTA0-}) in an independent manner by using the
Heisenberg-Euler (HE) effective Lagrangian
\cite{heisenberg:1936,weisskopf:1936}. This is the LO in the
derivative expansion of the effective action (\ref{DELTA_S}) but
contains all orders in the intensity. It has the well-known
proper-time representation \cite{schwinger:1951} (we follow the
nice review \cite{dunne:2004})
\bea
  \fl  \DL = \! - \frac{1}{8 \pi^2} \int\limits_0^\infty
  \frac{ds}{s} \, e^{-seE_c} \bigg\{\frac{e^2 \mathfrak{a}\mathfrak{b}
  s^2}{\tanh(e\mathfrak{b}s)   \tan(e\mathfrak{a}s)}
   - 1 - \frac{e^2s^2}{3}
  (\mathfrak{a}^2 - \mathfrak{b}^2) \bigg\} , \label{LHE}
\eea
and is exact for constant fields but remains approximately valid
for photon frequencies small compared to the electron mass as in
(\ref{NU}). The quantities $\pm i\mathfrak{a}$ and $\pm
\mathfrak{b}$ are the eigenvalues of the constant matrix
\be \label{GMUNU}
  G_{\mu\nu} \equiv F_{\mu\nu} + f_{\mu\nu} \; ,
\ee
consisting of both background and probe field. They are related to
the standard scalar and pseudoscalar invariants defined in
(\ref{S0}) and (\ref{PS0}) via
\bea
  \mathfrak{a}^2 - \mathfrak{b}^2 &=& -\frac{1}{2} G_{\mu\nu}G^{\mu\nu} \equiv 2 \SCS \; ,
  \label{S}  \\
  \mathfrak{a} \mathfrak{b} &=& - \frac{1}{4} G_{\mu\nu} \tilde{G}^{\mu\nu} \equiv \SCP
  \ ; \label{PS}
\eea
Note that in this appendix the invariants $\SCS$ and $\SCP$ denote
the contribution of \emph{both} background and fluctuation, cf.\
(\ref{GMUNU}), and thus are nonvanishing. The representation
(\ref{LHE}) contains all orders in the field $G_{\mu\nu}$. Low
intensities allow for a weak-field expansion the first two orders
of which are
\bea
  \DL = \frac{\alpha^2}{m_e^4} \left( \frac{8}{45} \SCS^2 +
  \frac{14}{45} \SCP^2  \right)
  + \frac{\alpha^3}{m_e^8} \left( \frac{256\pi}{315} \SCS^3 +
  \frac{416\pi}{315} \SCS \SCP^2 \right) \; . \label{LHE_NLO}
\eea
It is worth to point out that exactly the same numerical
coefficients appear as in (\ref{DELTA_PM}).

For what follows it is useful to write the LO of (\ref{LHE_NLO})
as
\be \label{LHE_LO}
  \DL = \frac{1}{2} \gamma_- \SCS^2 + \frac{1}{2} \gamma_+ \SCP^2 + O(\alpha^3) \; ,
\ee
with the couplings $\gamma_\pm$ given by
\be \label{GAMMA_PM}
  \gamma_+ \equiv 7 \xi \; , \quad \gamma_- \equiv 4 \xi \; ,
  \quad \xi \equiv \frac{\alpha}{45 \pi} \frac{1}{E_c^2} \; .
\ee
To approximate the polarisation tensor $\Pi^{\mu\nu}$  we start
from (\ref{LHE_NLO}), decompose into background $F$ and probe $f$
according to (\ref{GMUNU}) and expand the HE action, $\delta S =
\int d^4 x \, \DL$, to second order in the probe field $f$,
\be
  \delta S [F,f] = \sfrac{1}{8} \left( f_{\alpha \beta} ,
  \Omega ^{\alpha\beta\mu\nu} \, f_{\mu\nu}  \right) \; .
\ee
The tensor $\Omega ^{\alpha\beta\mu\nu}$ is proportional to the
second derivative of $\DL$,
\be \label{HE_DERIV}
  \Omega ^{\alpha\beta\mu\nu} \equiv 4 \left. \frac{\partial^2
  (\DL)}{\partial f_{\alpha \beta} \, \partial f_{\mu\nu}}
  \right|_{f=0} \; .
\ee
So far we have not exploited the fact that our background consists
of crossed fields. Note that the derivative in (\ref{HE_DERIV}) is
evaluated right at the background. It is easy to see that enormous
simplifications arise in the crossed-field case due to the
vanishing of the background invariants. The generic term in the HE
Lagrangian is of the form $\SCS^n \SCP^{2m}$ (as odd powers of
$\SCP$ are forbidden by CP invariance) with $n$ and $m$ integers.
Taking the two derivatives in (\ref{HE_DERIV}) at $f=0$ one will
always end up with (vanishing) powers of $\SCS$ and $\SCP$ unless
$n$ and $m$ are sufficiently small. The only surviving cases turn
out to be the Maxwell term, $(n=1$, $m=0$), and the LO
(\ref{LHE_LO}) with $n=2$, $m=0$ and $n=0$, $m=1$. Thus, for
crossed fields, the effective Lagrangian (\ref{LHE}) gets
truncated after the LO $\alpha^2/m_e^4$. In other words, the LO
describes the \textit{exact} dependence on intensity for
crossed-field background\footnote{An analogous observation has
been made for photon splitting in a plane-wave field
\cite{affleck:1987}.}! In view of these considerations the tensor
$\Omega^{\mu\nu\alpha\beta}$ simplifies to
\be \label{OMEGA}
  \Omega^{\mu\nu\alpha\beta} = \gamma_- F^{\alpha\beta}
  F^{\mu\nu} + \gamma_+ \tilde{F}^{\alpha\beta}
  \tilde{F}^{\mu\nu} \; .
\ee
Note that it is symmetric upon exchanging $(\mu\nu)
\leftrightarrow (\alpha\beta)$ and antisymmetric both in $\mu$ and
$\nu$ as well as $\alpha$ and $\beta$.

In momentum space the polarisation tensor $\Pi^{\mu\nu}$ is then
given by contracting (\ref{OMEGA}) twice with the wave vector $k$
associated with the probe field $f_{\mu\nu}$,
\be \label{PI1}
  \Pi^{\mu \nu} (k) \equiv \Omega^{\mu\alpha\nu\beta} k_\alpha
  k_\beta = \gamma_- \, b^\mu b^\nu +
  \gamma_+ \, \tilde{b}^\mu \tilde{b}^\nu
  \; .
\ee
To LO in $k^2$ this coincides with (\ref{PI_SPEC2}) as it should,
the nonvanishing eigenvalues being given by $\Pi_\pm = \gamma_\pm
b^2 (k)$. They imply the dispersion relations,
\be \label{DISP1}
  k^2 - \gamma_\pm b^2 (k) = 0 \;
\ee
and effective metrics
\be \label{DISP2}
   h_\pm^{\mu\nu} = g^{\mu\nu} + \gamma_\pm T^{\mu\nu} \; .
\ee
Introducing the index $n$ of refraction according to (\ref{INDEX})
the dispersion relations (\ref{DISP1}) and (\ref{DISP2}) become
four quadratic equations for $n$. Demanding $n=1$ for $\gamma_\pm
= 0$ singles out two of them. These are conveniently written in
terms of the abbreviations (\ref{T00}-\ref{HK}),
\be \label{N1}
  \fl n_\pm = \frac{1}{1 - \gamma_\pm (\mathcal{H} - \mathcal{H}_k)}
  \left\{ \sqrt{1 + \gamma_\pm \mathcal{H}_k - \gamma_\pm^2 \left[
  \mathcal{H}^2 - (\vcb{k} \cdot \vc{S})^2 - \mathcal{H}
  \mathcal{H}_k \right]} - \gamma_\pm  \vcb{k} \cdot \vc{S}
  \right\} \;
\ee
and have the  small-$\gamma_\pm$ expansion,
\bea \label{N2}
  \fl n_\pm &=& 1 + \gamma_\pm \left( \mathcal{H} - \vcb{k} \cdot \vc{S} -
  \mathcal{H}_k /2 \right) +
  \frac{1}{2} \gamma_\pm^2 \left[ (\mathcal{H} - \vcb{k} \cdot \vc{S} -
  \mathcal{H}_k)^2 -  \mathcal{H}_k /2 \right] + \ldots \; .
\eea
For a head-on collision of probe and background (\ref{N1}) yields
the simple expression
\be \label{N3}
  n_\pm = \frac{1 + \gamma_\pm \, I}{1 - \gamma_\pm \, I} = 1 +
  \frac{2 \gamma_\pm I}{1 - \gamma_\pm I} \equiv 1 + \Delta_\pm \; ,
\ee
which is exact to all orders in the intensity $I$. Noting that
\be
  \gamma_\pm I = \frac{11 \pm 3}{2} \frac{\alpha\epsilon^2}{45\pi}
  \; ,
\ee
the result (\ref{N3}) coincides with (\ref{DELTA0+}) and
(\ref{DELTA0-}).

\section*{References}

%\bibliographystyle{../../bibfiles/h-physrev}
%\bibliographystyle{JHEP}
%\bibliographystyle{unsrt}
%\bibliography{../../bibfiles/laser}

%\begin{thebibliography}{10}
%\end{thebibliography}

\end{document}